\documentclass[reprint,amsmath,showpacs,amssymb,aip,groupedaddress,pop]{revtex4-1}

\usepackage[english]{babel}
\usepackage{graphicx}
\usepackage{dcolumn}
\usepackage{bm}
\usepackage{hyperref}

\begin{document}

\title{Ab initio calculation of thermodynamic, transport, and optical properties of CH$_2$ plastics}

\author{D.~V.~Knyazev$^{1,2,3}$ and P.~R.~Levashov$^{1,4}$}
\affiliation{$^1$Joint Institute for High Temperatures RAS, Izhorskaya 13 bldg. 2, Moscow 125412, Russia \\ 
$^2$Moscow Institute of Physics and Technology (State University), Institutskiy per. 9, Dolgoprudny, Moscow Region, 141700, Russia,\\
$^3$State Scientific Center of the Russian Federation -- Institute for Theoretical and Experimental Physics of National Research Centre "Kurchatov Institute", Bolshaya Cheremushkinskaya 25, 117218, Moscow, Russia\\
$^4$Tomsk State University, Lenin Prospekt 36, Tomsk 634050, Russia}

\begin{abstract}

This work covers an \textit{ab initio} calculation of thermodynamic, transport, and optical properties of  plastics of the effective composition CH$_2$ at density 0.954 g/cm$^3$ in the temperature range from 5~kK up to 100~kK. The calculation is based on the quantum molecular dynamics, density functional theory and the Kubo-Greenwood formula. The temperature dependence of the static electrical conductivity $\sigma_{1_\mathrm{DC}}(T)$ has a step-like shape: $\sigma_{1_\mathrm{DC}}(T)$ grows rapidly for 5~kK~$\leq T\leq10$~kK and is almost constant for 20~kK~$\leq T\leq60$~kK. The additional analysis based on the investigation of the electron density of states (DOS) is performed. The rapid growth of $\sigma_{1_\mathrm{DC}}(T)$ at 5~kK~$\leq T\leq10$~kK is connected with the increase of DOS at the electron energy equal to the chemical potential $\epsilon=\mu$. The frequency dependence of the dynamic electrical conductivity $\sigma_1(\omega)$ at 5~kK has the distinct non-Drude shape with the peak at $\omega\approx10$~eV. This behavior of $\sigma_1(\omega)$ was explained by the dip at the electron DOS.

\end{abstract}

\maketitle

\section{INTRODUCTION}

The carbon-hydrogen plastics are widely used in the experiments in high energy density physics. These materials are often applied as ablators in experiments on the inertial confinement fusion \cite{Hammel:PhysPlasmas:2011, Wilson:PhysPlasmas:2011} (ICF) and laser-driven shock-wave measurements \cite{Boehly:PhysPlasmas:2004, Hicks:PRB:2009}. The contrast improvement of intense laser pulses is another fruitful application of plastics \cite{Povarnitsyn:LasPartBeams:2013}.

Intense laser pulses ($I\sim10^{19}\div10^{20}$~W/cm$^2$) are preceded by prepulses, caused by the amplified spontaneous emission. The intensity of the prepulse may be many orders of magnitude (say, $10^8$) lower than the intensity of the main pulse. However, for the huge intensities of the main pulse ($10^{20}$~W/cm$^2$), the prepulse is intense enough ($10^{12}$~W/cm$^2$) to damage the target significantly \cite{McKenna:PTRSA:2006}. One of the possible remedies is to place a film in front of the target \cite{Povarnitsyn:PhysPlasmas:2012}. The prepulse creates plasma from the film, that reflects the prepulse radiation and, thus, protects the target. Then the plasma produced from the film expands and, if the film thickness was chosen correctly, becomes transparent right before the arrival of the main pulse. The usage of the plastic film is particularly promising \cite{Povarnitsyn:LasPartBeams:2013}, because it yields lower fluxes of re-emitted X-rays which may also damage the target.

The numerical simulation is necessary to choose the correct thickness of the film. The intensity of the prepulse is low enough ($10^{12}$~W/cm$^2$), so that the two-temperature hydrodynamic simulation may be applied to describe its action \cite{Povarnitsyn:PhysPlasmas:2012, Povarnitsyn:LasPartBeams:2013}. The two-temperature hydrodynamic model \cite{Povarnitsyn:ASS:2012, Inogamov:JOT:2014} requires the knowledge of the matter properties: two-temperature equation of state, optical characteristics (complex dielectric function), thermal conductivity, electron-ion coupling. The necessity of this information stimulates experimental and theoretical research on matter properties. In this work we are focused on the theoretical investigation of thermodynamic, transport and optical properties of plastics.

The matter of the film undergoes complicated evolution under the action of the prepulse \cite{Povarnitsyn:PhysPlasmas:2012}. At first the matter is heated almost isochorically, the two-temperature effects are of significant importance at this stage. Then the matter is compressed by the induced shock waves and the temperature grows even further. After that the matter expands and the temperature drops.

Different methods of simulation are applicable to obtain matter properties in different regions of phase diagram. The average atom model \cite{Ovechkin:HEDP:2014, Sinko:HEDP:2013} may be used to calculate matter properties at high temperatures and compressions. The model may be applied both for one-component \cite{Novikov:MMCS:2011} and multi-component systems \cite{Faussurier:PhysPlasmas:2014}. The chemical plasma model may be employed to calculate matter properties at low densities \cite{Redmer:PRE:1999, Apfelbaum:CPP:2013}. The quantum molecular dynamics (QMD) simulation in conjunction with the Kubo-Greenwood formula is well suited to calculate matter properties during the near-isochoric heating. In this work we calculate the properties of plastics along the normal isochor via the latter approach.

The QMD simulation based on the density functional theory (DFT) was used in numerous works for the calculation of thermodynamic properties, including Hugoniots \cite{Desjarlais:PRB:2003, Knudson:PRL:2009, Minakov:JAP:2014} and equations of state \cite{Holst:PRB:2008, French:PRB:2009} (EOS).

The QMD simulation together with the Kubo-Greenwood formula is also widely used to calculate transport and optical properties. This method was first applied in early works \cite{Fois:JPC:1988, Silvestrelli:PRB:1999} and then in well-known papers \cite{Collins:PRB:2001, Desjarlais:PRE:2002, Recoules:PRB:2005}. Since that time the Kubo-Greenwood formula was used to calculate transport and optical properties of many systems \cite{Laudernet:PRB:2004, French:PhysPlasmas:2011, Lambert:PhysPlasmas:2011, Norman:CPP:2013, Norman:PRE:2015}. In our previous works \cite{Povarnitsyn:CPP:2012, Knyazev:COMMAT:2013, Knyazev:PhysPlasmas:2014} we have employed this technique to calculate transport and optical properties of aluminum.

The carbon-hydrogen plastics have been also recently explored by the QMD technique. The research was mainly focused on the calculation of the principal Hugoniots and the properties of plastics in the compressed state. The attention was also scattered among the plastics of various chemical compositions.

The plastic of the effective composition C$_2$H$_3$ was investigated for ICF applications \cite{Lambert:PRE:2012}. Thermodynamic and transport properties (including thermal conductivity) were calculated at high temperatures from 5 to 40~eV and high densities from 7 to 9~g/cm$^3$.

The plastic of the effective composition CH$_{1.36}$ was explored recently \cite{Hamel:PRB:2012} with the same purpose. The EOS was calculated for the wide range of densities from 1.8 to 10~g/cm$^3$ and for the temperatures from 4 to 100~kK.

Polystyrene (effective composition CH) was explored in the papers \cite{Wang:PhysPlasmas:2011, Chantawansri:JCP:2012, Hu:PRE:2014:1}. The work \cite{Wang:PhysPlasmas:2011} contains the calculation of the principal Hugoniot for the compressions from 2 to 3.2; the information on the static electrical conductivity and reflectivity is also available. The principal Hugoniot for the compressions from 1.25 to 2 is presented in the paper \cite{Chantawansri:JCP:2012}. In work \cite{Hu:PRE:2014:1}, the principal Hugoniot for the compressions from 1.85 to 3.9 is presented together with the reflectivity data.

Polyethylene (effective composition CH$_2$) and poly(4-methyl-1-pentene) (PMP, effective composition CH$_2$) were investigated in the works \cite{Mattsson:PRB:2010, Chantawansri:JCP:2012, Horner:PRB:2010}. The principal Hugoniot of polyethylene for compressions from 1.15 to 2.1, the principal Hugoniot of PMP for the compressions from 1.4 to 3.2 were calculated in the paper \cite{Mattsson:PRB:2010}; the work also contains the analysis of polyethylene structural properties under elevated temperatures and densities. The work \cite{Chantawansri:JCP:2012} covers the calculation of the principal Hugoniot of polyethylene for the compressions from 1.25 to 1.7; the structural properties of polyethylene are also investigated. The paper \cite{Horner:PRB:2010} contains the results on thermodynamic, transport and optical properties of plastics of the effective composition CH$_2$ for the densities from 1 to 3~g/cm$^3$ and temperatures from 1 to 4~eV; the influence of aluminum impurities on the matter properties is investigated.

The properties of plastics of the effective composition CH$_2$ at the normal isochor 0.954~g/cm$^3$ are necessary for the simulation of prepulse action on the plastic film \cite{Povarnitsyn:LasPartBeams:2013}. This composition and normal density correspond to the high-density polyethylene. The review of the previous works available above demonstrates the lack of the results on the plastic properties at the normal density. This motivated us to perform the current work.

Our paper is organized as follows. We briefly discuss the calculation technique in Sec.~\ref{Sec:Technique}. Then we consider the technical parameters used during the calculation, including the creation of the initial configuration (Sec.~\ref{Sec:Parameters}). Section~\ref{Sec:Results} contains obtained results on thermodynamic, transport and optical properties of CH$_2$ plastics at the density 0.954~g/cm$^3$ and temperatures from 5~kK up to 100~kK. Then we analyse the results obtained (Sec.~\ref{Sec:Discussion}). The comparison with the work \cite{Horner:PRB:2010} is performed; the results on the static electrical conductivity are explained by the investigation of electron DOS.

\section{CALCULATION TECHNIQUE}
\label{Sec:Technique}

Here we will present only the brief description of the computation technique. It was described in detail in our previous paper \cite{Knyazev:COMMAT:2013} and earlier works \cite{Collins:PRB:2001, Desjarlais:PRE:2002, Recoules:PRB:2005, Holst:PRB:2011}.

Three main stages are included to the calculation: QMD simulation and calculation of the thermodynamic properties; precise resolution of the band structure; calculation of the transport and optical properties according to the Kubo-Greenwood formula.

At the first stage the atoms of carbon and hydrogen in the necessary proportion are placed to the supercell with periodic boundary conditions. The volume of the supercell is chosen to yield the correct density of plastic. At the initial moment the ions are placed to the auxiliary simple cubic lattice (see Sec.~\ref{Sec:Parameters} for more detail); the QMD simulation is used to calculate ionic trajectories in the further moments. At every step of QMD simulation the electronic structure is calculated by the DFT approach via the solving of the Kohn-Sham equations. Ions are treated classically, the forces acting on the ions are calculated at every step via the Hellmann-Feynman theorem. The temperature of electrons is determined by the parameter in the Fermi-Dirac distribution; the temperature of ions is maintained by the Nos\'e-Hoover thermostat. The QMD simulation is performed using the Vienna \textit{ab initio} simulation package (VASP) \cite{Kresse:PRB:1993, Kresse:PRB:1994, Kresse:PRB:1996}.

The VASP code provides the temporal dependences of the following properties: total energy of electrons and ions without kinetic energy of ions $E(t)-E^\mathrm{kin}_i(t)$, kinetic energy of ions $E^\mathrm{kin}_i(t)$, total pressure of electrons and ions without kinetic contribution of ions $p(t)-p^\mathrm{kin}_i(t)$, kinetic contribution of ions $p^\mathrm{kin}_i(t)$. When the simulation starts, these properties vary until equilibrium (or at least metastable, see Sec.~\ref{Sec:Parameters}) section of MD run is reached. At the equilibrium section of evolution the properties mentioned fluctuate around their equilibrium values. To obtain thermodynamic properties the temporal dependences are averaged over the equilibrium section. Every step of the QMD simulation from the equilibrium section is used for averaging. Thus we obtain thermodynamic values $E-E^\mathrm{kin}_i$, $E^\mathrm{kin}_i$, $p-p^\mathrm{kin}_i$, $p^\mathrm{kin}_i$. Then we may calculate the total energy of electrons and ions $E= (E-E^\mathrm{kin}_i)+E^\mathrm{kin}_i$ and the total pressure of electrons and ions $p=(p-p^\mathrm{kin}_i)+p^\mathrm{kin}_i$. For the sake of convenience the energy obtained is divided by the total mass of the supercell, thus yielding the specific energy.

Thus we obtain the temperature dependence of energy $E(T)$, calculated at separate temperature points along the isochor. We compute the heat capacity $C_v(T)$ performing numerical differentiation $dE/dT$. The derivative is calculated as a half-sum of forward and backward finite-differences of the first-order accuracy. We compute both $dE/dT$ and $d(E-E^\mathrm{kin}_i)/dT$.

At the second stage the precise resolution of the band structure is performed. We use much less number of ionic configurations for the calculation of the transport properties, than for the calculation of the thermodynamic properties. So we select only several ionic configurations from the equilibrium section of the MD run. For these selected configurations we calculate the electronic structure. Again, as well as during the QMD simulation, the Kohn-Sham equations are solved. Though now we may use a larger cut-off energy, number of bands, number of \textbf{k}-points in the Brillouin zone. The higher values of these parameters increase the precision of calculation. At this stage we obtain energy eigenvalues, corresponding wave functions and occupation numbers; this information is used at the third stage to calculate the dynamic Onsager coefficients. Precise resolution of the band structure is performed using the VASP package.

At the third stage the dynamic Onsager coefficients $L_{mn}(\omega),~m,~n=1,2$ are calculated for each ionic configuration according to the Kubo-Greenwood formula:
\begin{multline}
L_{mn}(\omega)=(-1)^{m+n}\frac{1}{e^{m-1}(eT)^{n-1}} \frac{2\pi e^2\hbar^2}{3m_e^2\omega\Omega}\times\\
\times\sum_{i,j,\alpha,\mathbf{k}}W(\mathbf{k})\left(\frac{\epsilon_{i,\mathbf{k}}+\epsilon_{j,\mathbf{k}}}{2}-\mu\right)^{m+n-2}\times\\
\times\left|\left\langle\Psi_{i,\mathbf{k}}\left|\nabla_\alpha\right|\Psi_{j,\mathbf{k}}\right\rangle\right|^2\left(f(\epsilon_{i,\mathbf{k}})-f(\epsilon_{j,\mathbf{k}})\right)\times\\\times\delta(\epsilon_{j,\mathbf{k}}-\epsilon_{i,\mathbf{k}}-\hbar\omega).
\label{Eq:DynamicOnsager}
\end{multline}
Here $\Psi_{i,\mathbf{k}}$ and $\epsilon_{i,\mathbf{k}}$ are the electronic wave function and energy eigenvalue respectively, corresponding to the particular band $i$ and the point in the Brillouin zone $\mathbf{k}$. $f(\epsilon_{i,\mathbf{k}})$ is the Fermi-weight of the particular band, $W(\mathbf{k})$---the weight of the particular \textbf{k}-point. $\mu$ is the chemical potential, $\omega$---the frequency of the applied electric field, $\Omega$---the volume of the supercell, $\hbar$ is the reduced Planck constant, $m_e$ is the electron mass, $T$ stands for the temperature of electrons and ions here.

The delta-function in the Kubo-Greenwood formula (\ref{Eq:DynamicOnsager}) is broadened \cite{Desjarlais:PRE:2002} by the the Gaussian function with the standard deviation $\Delta E$. The intuitively clear derivation of the Kubo-Greenwood formula is present in paper \cite{Moseley:AJP:1978}, the derivation of the form (\ref{Eq:DynamicOnsager}) with half-sum $\frac{\epsilon_{i,\mathbf{k}}+\epsilon_{j,\mathbf{k}}}{2}$ may be found in the work \cite{Holst:PRB:2011}, additional discussion of the advantage of the formula with the half-sum  is present in our previous work \cite{Knyazev:COMMAT:2013}.

We use the VASP module optics.f90 to calculate matrix elements $\left\langle\Psi_{i,\mathbf{k}}\left|\nabla_\alpha\right|\Psi_{j,\mathbf{k}}\right\rangle$. The optics.f90 module provides the correct calculation of the matrix elements with necessary corrections due to the usage of the projector augmented-wave (PAW) pseudopotentials taken into account. We have also created special parallel program module that uses matrix elements obtained from VASP to calculate the dynamic Onsager coefficients according to the Kubo-Greenwood formula (\ref{Eq:DynamicOnsager}).

The dynamic Onsager coefficients calculated for different ionic configurations are averaged. The real part of the dynamic electrical conductivity $\sigma_1(\omega)$ is simply the $L_{11}(\omega)$ Onsager coefficient.

No physical meaning is assigned to the other dynamic Onsager coefficients. They are necessary only to obtain the static Onsager coefficients $L_{mn},m,n=1,2$ via the extrapolation to the zero frequency: $L_{mn}=\lim_{\omega\to0}L_{mn}(\omega)$. The static electrical conductivity $\sigma_{1_\mathrm{DC}}\equiv L_{11}$; the thermal conductivity $K$ is expressed via the static Onsager coefficients as follows:
\begin{equation}
K=L_{22}-\frac{L_{12}L_{21}}{L_{11}}.
\label{Eq:K_exact}
\end{equation}
It should be mentioned, that $K$ contains the additional second term in Eq.~(\ref{Eq:K_exact}), called the \textit{thermoelectric term}. It was shown theoretically \cite{Ashcroft:1976} and confirmed numerically \cite{Recoules:PRB:2005, Knyazev:COMMAT:2013, Knyazev:PhysPlasmas:2014}, that for metals at relatively low temperatures the relative contribution of the thermoelectric term is rather small. In this work we check the relative contribution of the thermoelectric term to the thermal conductivity of plastics at different temperatures.

The experimentally discovered Wiedemann-Franz law is valid for metals at low temperatures:
\begin{equation}
\frac{K(T)}{\sigma_{1_\mathrm{DC}}(T)\cdot T}=\mathrm{const}=L_\mathrm{ideal~deg}=\frac{\pi^2}{3}\frac{k^2}{e^2}.
\label{Eq:Lorenz}
\end{equation}
The static electrical conductivity $\sigma_{1_\mathrm{DC}}(T)$ and the thermal conductivity $K(T)$ may depend on the temperature, whereas the ratio $L$, called the Lorenz number, is a constant. The ideal degenerate value $L_\mathrm{ideal~deg}=\frac{\pi^2}{3}\frac{k^2}{e^2}=2.44\times10^{-8}$~W$\cdot\Omega\cdot$K$^{-2}$ may be justified theoretically for metals at low temperatures \cite{Chester:PPS:1961, Ashcroft:1976}. If aluminum is exposed to elevated temperatures at first the Wiedemann-Franz law holds \cite{Recoules:PRB:2005, Knyazev:COMMAT:2013}, but at even higher temperatures it is violated \cite{Knyazev:PhysPlasmas:2014}. It is not evident, whether the Wiedemann-Franz law is valid for plastics. However, sometimes \cite{Horner:PRB:2010} the Wiedemann-Franz law is also used for plastics. In this work we obtain the static electrical conductivity and the thermal conductivity in the \textit{ab initio} calculation and check the Wiedemann-Franz law for plastics.

We have also calculated electron DOS. During the precise resolution of the band structure the information on the energy eigenvalues was obtained. The whole range of energy eigenvalues $\epsilon$ (hereafter called the electron energy for simplicity) was divided into the sections of $\Delta E_\mathrm{rect}$ width. The number of bands falling within each section was determined and divided by the width of the section. The obtained value $g(\epsilon)$ was assigned to the electron energy at the middle of the section. Each band was counted only once, i.e. the spin degeneracy was not taken into account. The results presented in this paper contain the values $g(\epsilon)/\Omega$, where $\Omega$ is the volume of the supercell. For the given material at the fixed density and temperature for each given energy $\epsilon$ $g(\epsilon)/\Omega$ remains almost (neglecting small fluctuations) fixed as the volume $\Omega$ is varied. If we present results in the form $g(\epsilon)/\Omega$ we may compare the DOS calculated for different numbers of atoms in the supercell. We could definitely use methods based on the Gaussian smearing yielding the smoother curves $g(\epsilon)/\Omega$; however, in this work we use the simple method described above which is enough for the analysis.

\section{CALCULATION PARAMETERS}
\label{Sec:Parameters}

The bulk of our calculations was performed for 120 atoms (40 atoms of carbon, 80 atoms of hydrogen) in the computational supercell. Only several additional runs were made for 249 atoms to check the dependence on the number of atoms.

We have chosen the QMD step to be 0.2~fs. Such a small value ensures the stable work of the VASP package for different initial ionic configurations. At larger values of step, for some initial ionic configurations the kinetic energy of ions grows rapidly and the system becomes numerically unstable.

The Nos\'e mass parameter SMASS is also an important technical parameter. We have set SMASS to 0, thus we let VASP to set SMASS automatically. Thus the SMASS parameter varied from 0.05 at 5~kK to 0.98 at 100~kK. If the SMASS parameter is too large, the kinetic energy of ions has a large period of fluctuations and it also takes significantly more time steps for the kinetic energy of ions to equilibrate.

The initial ionic configuration should also be chosen with care. If we wanted to simulate a plastic in a solid phase we would have to put the ions in the realistic ionic configuration; the long polymeric chains would have to be taken into account. This would also require a large computational cell, and therefore huge computational efforts. We are interested in the properties of plastic at higher temperatures, when all the polymeric chains and even chemical bonds are destroyed. Thus we could put ions to the random positions at the initial moment and expect, that QMD simulation would produce correct ionic configurations during the simulation. However, if the ions are placed randomly, they may come arbitrarily close to each other. This leads to the rapid growth of the kinetic energy at the start of simulation and may cause the numerical instability at the large steps of simulation. To avoid these problems we construct an auxiliary simple cubic lattice and put ions of carbon and hydrogen to the random positions at the nodes of this auxiliary lattice. Thus we restrict the lowest possible distance between the ions. If we choose the initial ionic configuration this way, the kinetic energy of ions grows more slowly  at the start of simulation; this makes the time necessary for equilibration of kinetic energy of ions lower. It also provides opportunity to increase the step of simulation (though we keep it low at 0.2 fs). 

At the start of simulation we put ions to the random positions at the nodes of the auxiliary simple cubic lattice. Then we perform 15000 steps of simulation. Given the simulation step is 0.2 fs, we track 3 ps of the system evolution.

Following the procedure of configuration generation described above, we need to know only the relative fractions of carbon and hydrogen. Further in this paper we call this unordered state a \textit{plastic of the effective composition CH$_\textit{2}$}.

The setting of the initial ionic configuration of plastics for the QMD simulation is an involved problem; the methods of its construction vary from quite simple to rather complicated.

In some works \cite{Lambert:PRE:2012, Hu:PRE:2014:1} the initial configuration is not specified and only the average composition of the supercell is mentioned.

In other works \cite{Horner:PRB:2010, Wang:PhysPlasmas:2011} the supercell is filled with some small molecules. CH$_2$ molecules are placed to the random positions to set the initial configuration for polyethylene in the paper \cite{Horner:PRB:2010}; the problems of the molecules coming arbitrarily close to each other (similar to our problems described above) are encountered. In Ref.~\cite{Wang:PhysPlasmas:2011} the supercell is filled with the C$_8$H$_8$ units to set the initial configuration for polystyrene.

The crystalline polyethylene structure may be obtained by placing ions to the polymeric chains inside the tetragonal supercell with periodic boundary conditions \cite{Mattsson:PRB:2010}. Such an approach has a drawback: this tetragonal phase of polyethylene has not been found experimentally \cite{Chantawansri:JCP:2012}.

The most complicated approach is to use a polymer builder program to construct some number of polymeric chains \cite{Chantawansri:JCP:2012, Hamel:PRB:2012}. Then the supercell is exposed to several cycles of energy minimization and annealing to obtain the structure with the lowest energy \cite{Chantawansri:JCP:2012}.

In this work we use the simplest procedure described above to set the initial ionic configuration. We can not assert, that we reach a thermodynamically equilibrium state. But at least we can be sure that we explore a state, that is metastable during the simulation time. The previous works \cite{Mattsson:PRB:2010, Chantawansri:JCP:2012} show, that the time of several picoseconds is quite sufficient for chemical bonds to break and form. This ensures that the simple procedure employed in this work is good enough to reasonably reproduce the properties of plastics.

During the QMD simulation we have used the local density approximation (LDA) for exchange-correlation functional. The exchange part of the functional had a common $\sim\rho^{4/3}$ dependence, whereas the correlation part had the Ceperley-Alder parametrisation \cite{Perdew:PRB:1981}. We have used PAW pseudopotentials \cite{Blochl:PRB:1994, Kresse:PRB:1999}, 4 electrons per carbon atom and 1 electron per hydrogen one were included in the calculation. At the highest temperature of 100~kK we have also tried to include 6 electrons per carbon atom to the calculation. This only added 40 (2 electrons per band) fully occupied bands which were located approximately 250~eV lower than other bands. These fully occupied bands apparently do not affect the behavior of other electrons.

During the QMD simulation we have used the energy cut-off $E_\mathrm{cut}=300$~eV and 1 \textbf{k}-point ($\Gamma$-point) in the Brillouin zone. The number of bands increased as the temperature was increased; the number of bands was selected as described in our previous paper \cite{Knyazev:COMMAT:2013}. The bands with the occupation numbers not less than $5\cdot10^{-6}$ were considered occupied.

We have used configurations from 2500 to 12500 (corresponding to time from 0.5~ps to 2.5~ps) to average over the temporal dependences of energy and pressure. We have chosen 15 ionic configurations for the calculation of transport and optical properties: the first one was chosen at step 999 (0.2~ps); the period between neighboring configurations was 1000 time steps.

The precise resolution of the band structure was performed for each of the chosen ionic configurations. The exchange-correlation functional, the pseudopotential, the energy cut-off and the number of \textbf{k}-points during the precise resolution of the band structure were the same as during the QMD simulation. Only the number of bands was increased to enable the calculation of optical properties for the frequencies up to 40~eV. The number of bands was selected as described earlier \cite{Knyazev:COMMAT:2013}.

The dynamic Onsager coefficients were calculated for the frequencies from 0.005~eV up to 40~eV with the frequency step of 0.005~eV. The broadening of the $\delta$-function in the Kubo-Greenwood formula was $\Delta E=0.2$~eV. The simple linear extrapolation was used to obtain the static Onsager coefficients from the dynamic ones (see Ref.~\cite{Knyazev:COMMAT:2013} for details).

The width of the sections $\Delta E_\mathrm{rect}$ during the calculation of DOS should be selected the same way as the broadening $\Delta E$ in the Kubo-Greenwood formula (see detailed description in our previous paper \cite{Knyazev:COMMAT:2013}). $\Delta E_\mathrm{rect}$ should be small enough for the physical dependence on the DOS curves not to be smoothed; $\Delta E_\mathrm{rect}$ should not be too small, otherwise the oscillations will appear. In this work $\Delta E_\mathrm{rect}$ was chosen to be 2~eV both for calculations with 120 and 249 atoms in the supercell.

The comparison of our results with experimental works, reference data and calculations of other authors was performed for aluminum in our previous works \cite{Povarnitsyn:CPP:2012, Knyazev:COMMAT:2013}. Our calculations were permanently in good agreement with other results. This confirms that the obtained properties of plastics are also reliable. Moreover, additional comparison of our results on plastics with those of Horner \textit{et al.} \cite{Horner:PRB:2010} was performed in this work (Sec.~\ref{Sec:Discussion}).

\section{RESULTS}
\label{Sec:Results}

Thermodynamic, transport, and optical properties of plastics of the effective composition CH$_2$ were calculated for the density of 0.954~g/cm$^3$ and the temperatures from 5~kK up to 100~kK.

\begin{figure}
\includegraphics[width=0.95\columnwidth]{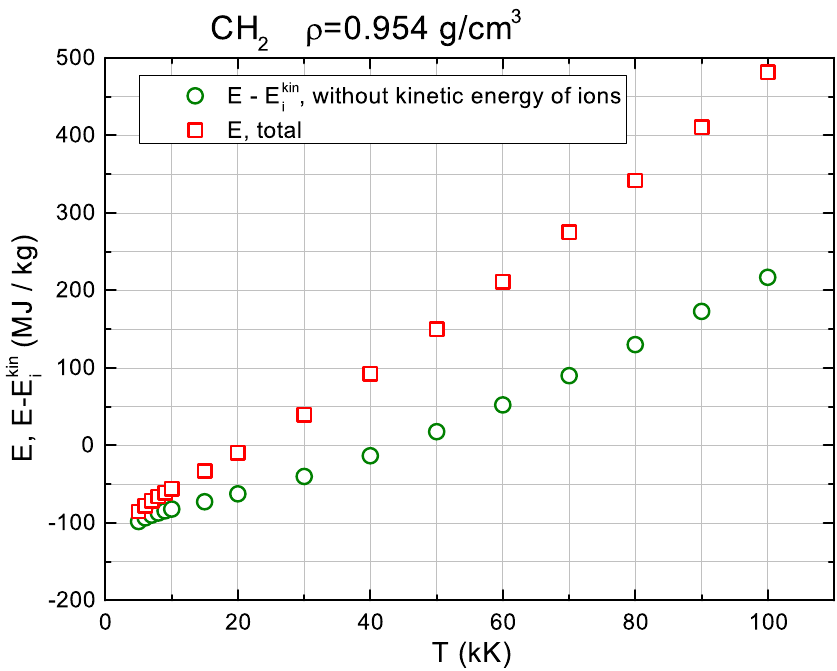}
\caption{Temperature dependences of specific energy of plastics. 120 atoms. Squares---kinetic contribution of ions is taken into account; circles---kinetic contribution of ions is neglected.}
\label{Fig:energy}
\end{figure}

\begin{figure}
a)~\includegraphics[width=0.95\columnwidth]{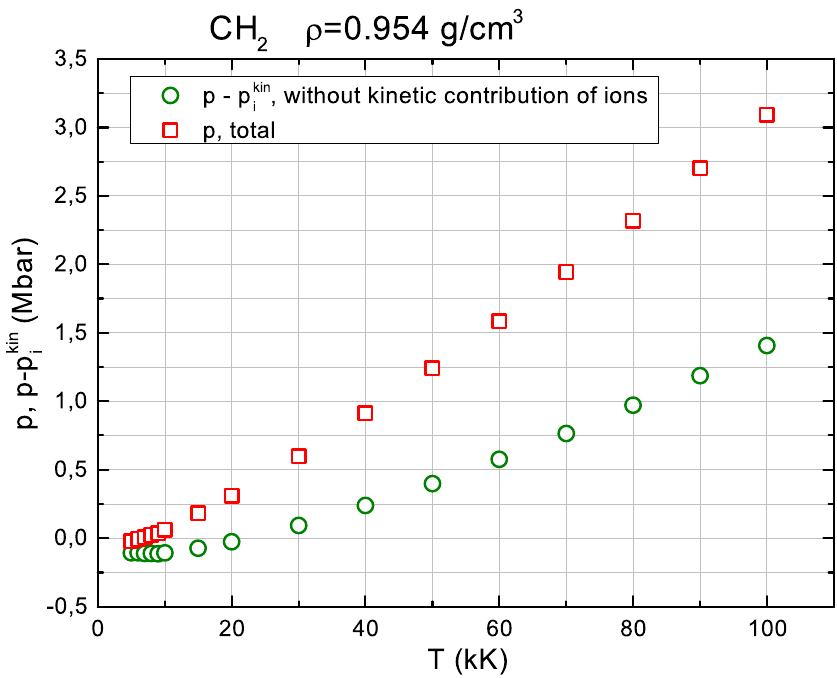}
b)~\includegraphics[width=0.95\columnwidth]{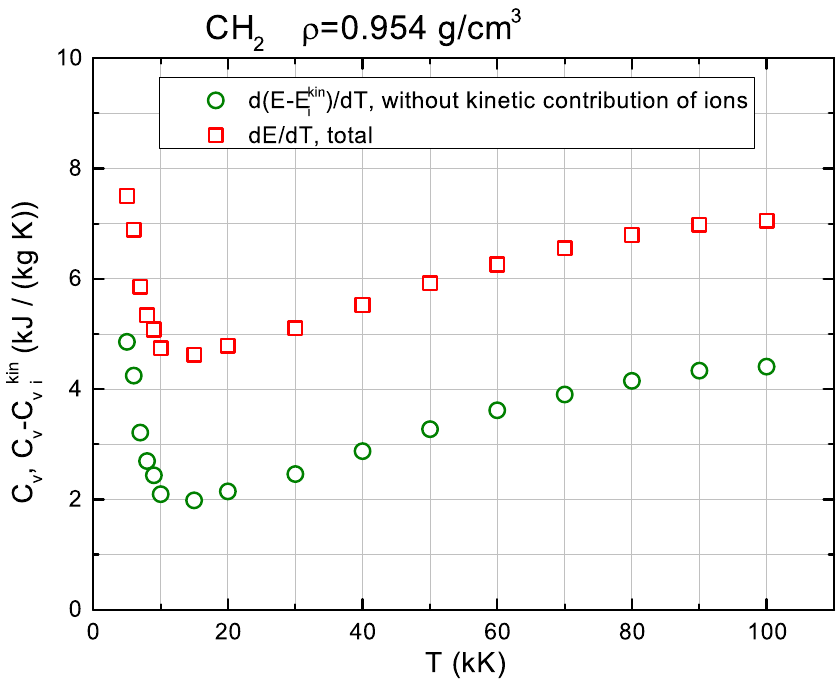}
\caption{Temperature dependences of thermodynamic properties of plastics. 120 atoms. Squares---kinetic contribution of ions is taken into account; circles---kinetic contribution of ions is neglected. (a) Pressure. (b) Specific heat capacity.}
\label{Fig:thermodynamic_der}
\end{figure}

The temperature dependences of thermodynamic properties of plastics are presented in Figs.~\ref{Fig:energy}-\ref{Fig:thermodynamic_der}.

Fig.~\ref{Fig:energy} shows the total energy of electrons and ions $E$ and the energy without kinetic energy of ions $E-E^\mathrm{kin}_i$. Both $E$ and $E-E^\mathrm{kin}_i$ grow as the temperature increases.

Fig.~\ref{Fig:thermodynamic_der}(a) shows the total pressure of electrons and ions $p$ and the pressure without the kinetic contribution of ions $p-p^\mathrm{kin}_i$. $p-p^\mathrm{kin}_i$ is negative for the temperatures $T\lesssim20$~kK. $p-p^\mathrm{kin}_i$ is almost constant (about $-114\div-106$~kbar) for the temperatures from 5~kK to 10~kK. $p-p^\mathrm{kin}_i$ grows at the increase of temperature. The kinetic contribution of ions $p^\mathrm{kin}_i$ is always positive and rises as the temperature grows. At low temperatures the positive contribution of $p^\mathrm{kin}_i$ somewhat compensates the negative one of $p-p^\mathrm{kin}_i$. The total pressure $p$ is negative at $T\lesssim 7$~kK ($-23$~kbar at 5~kK) but becomes positive at higher temperatures. Negative pressure $p$ at relatively low temperatures is probably caused by small errors in reproduction of normal density of elements and compounds by DFT \cite{Grabowski:PRB:2007}.

Fig.~\ref{Fig:thermodynamic_der}(b) shows the total specific heat capacity of electrons and ions $C_v$ and the heat capacity without kinetic contribution of ions $C_v-C_{v~i}^\mathrm{kin}$. The kinetic contribution of ions $C_{v~i}^\mathrm{kin}$ is the same for all temperatures. $C_v$ and $C_v-C_{v~i}^\mathrm{kin}$ demonstrate non-monotonic behavior: their values decrease at $T\leq15$~kK and increase at $T\geq15$~kK.

The results on energy and pressure were obtained for 120 atoms in the supercell and remained almost the same at several additional calculations performed with 249 atoms in the supercell.

\begin{figure}
a)~\includegraphics[width=0.95\columnwidth]{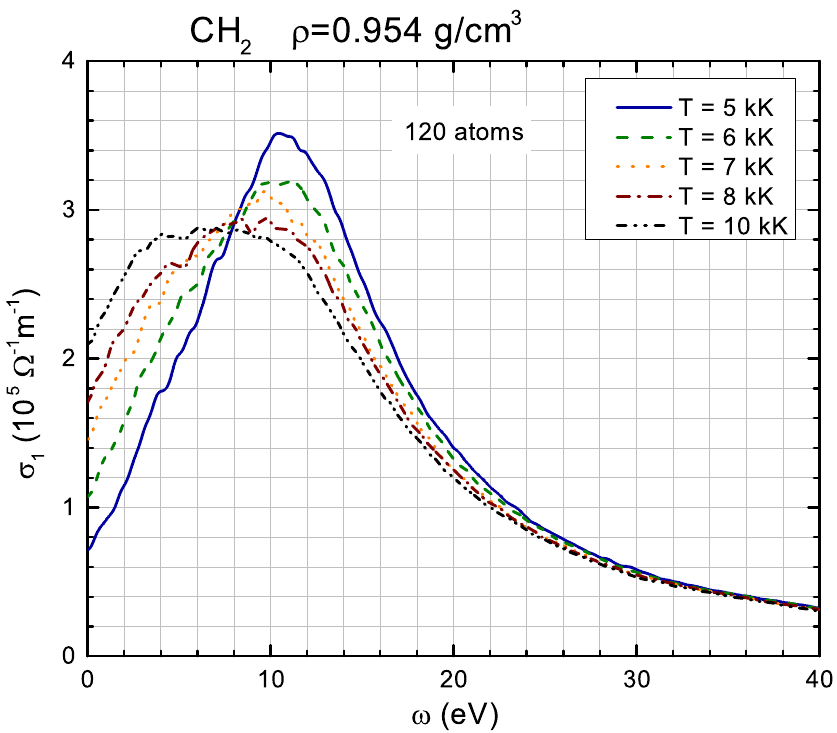}
b)~\includegraphics[width=0.95\columnwidth]{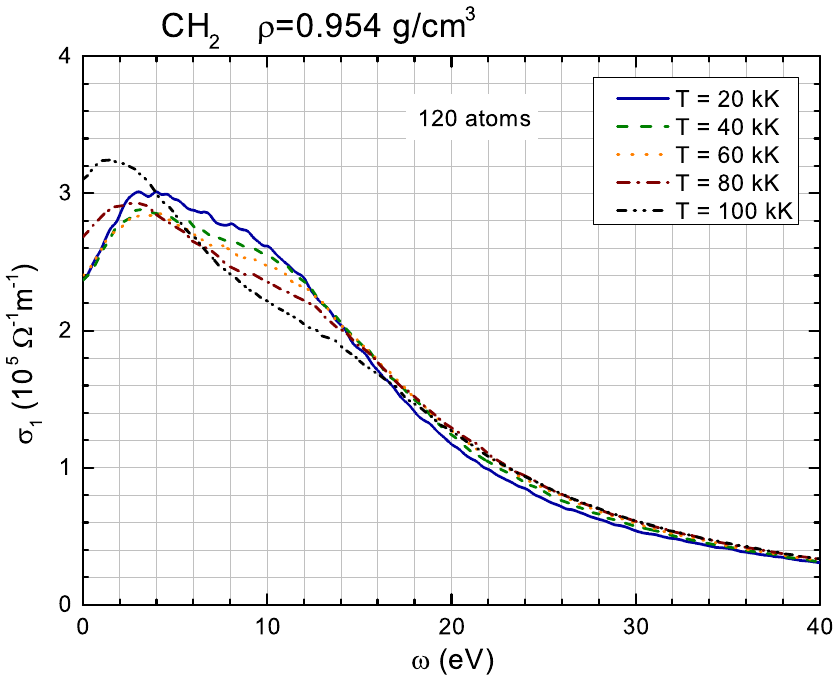}
\caption{Frequency dependences of the dynamic electrical conductivity of plastics for different temperatures. 120 atoms. (a) Temperature range $5\div10$~kK. (b) Temperature range $20\div100$~kK.}
\label{Fig:optical}
\end{figure}

The results on the dynamic electrical conductivity of plastics are presented in Fig.~\ref{Fig:optical}. 

Fig.~\ref{Fig:optical}(a) shows frequency dependences of the dynamic electrical conductivity of plastics for 5 temperatures from 5~kK to 10~kK. The curves have the distinct non-Drude shapes. At $T=5$~kK $\sigma_1(\omega)$ has the prominent peak at the frequency of about 10~eV. The curves change significantly as the temperature grows. At temperature rise the peak becomes smoother and moves to the lower frequencies. The dynamic conductivity grows at low frequencies as the temperature increases.

Fig.~\ref{Fig:optical}(b) shows frequency dependences of the dynamic electrical conductivity of plastics for 5 temperatures from 20~kK to 100~kK. For the temperature range from 20~kK to 60~kK the curves remain almost invariable if the temperature rises. This is especially noticeable if $\sigma_1(\omega)$ for 40~kK and 60~kK are compared. For the temperature range from 60~kK to 100~kK the curves begin to change slightly. The peak becomes sharper again and moves to even lower frequencies. The dynamic conductivity grows at low frequencies.

\begin{figure}
a)~\includegraphics[width=0.95\columnwidth]{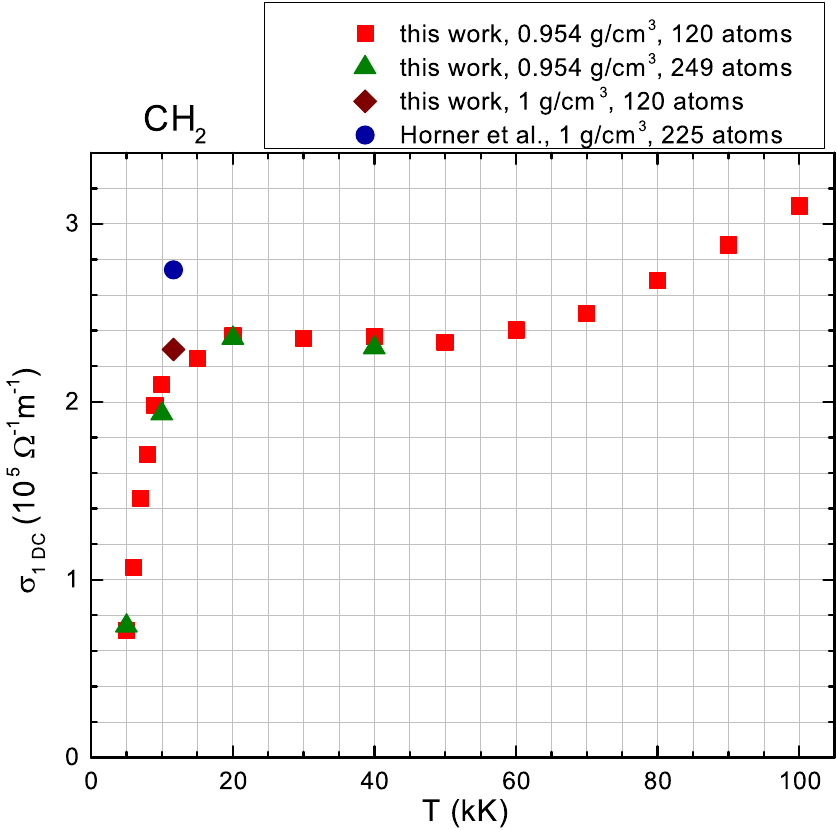}
b)~\includegraphics[width=0.95\columnwidth]{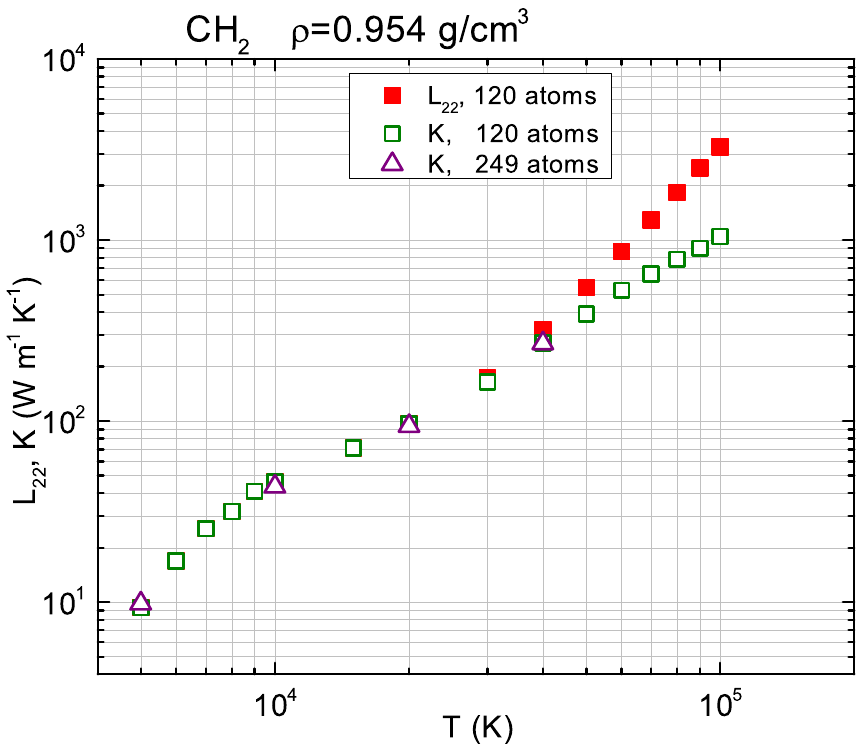}
\caption{Temperature dependences of the transport properties of CH$_2$ plastics. (a) Static electrical conductivity. Squares---$\rho=0.954$~g/cm$^3$, 120 atoms; triangles---$\rho=0.954$~g/cm$^3$, 249 atoms. Diamond---$\rho=1$~g/cm$^3$, $T=1$~eV, this work, 120 atoms; circle---$\rho=1$~g/cm$^3$, $T=1$~eV, Horner \textit{et al.}\cite{Horner:PRB:2010}. (b) Thermal conductivity, this work. Filled squares---the Onsager coefficient $L_{22}$ (thermoelectric term is neglected), 120 atoms; empty squares---thermal conductivity $K$ (thermoelectric term is taken into account), 120 atoms; empty triangles---thermal conductivity $K$, 249 atoms.}
\label{Fig:transport}
\end{figure}

The results on the transport properties of plastics are shown in Fig.~\ref{Fig:transport}.

Fig.~\ref{Fig:transport}(a) shows the temperature dependence of the static electrical conductivity. The results on the static conductivity show similar behavior as the results on the dynamic one shown above. The whole temperature range may be divided into three smaller sections: 1) 5~kK to 10~kK, the conductivity grows rapidly; 2) 20~kK to 60~kK, the conductivity is almost constant as the temperature is varied; 3) 60~kK to 100~kK, slow growth of the static conductivity.

Fig.~\ref{Fig:transport}(b) shows temperature dependences of the Onsager coefficient $L_{22}$ and the thermal conductivity $K$. The thermal conductivity grows monotonically in the whole range of temperatures under consideration. At low temperatures $K$ (thermal conductivity with thermoelectric term taken into account) coincides with $L_{22}$ (thermal conductivity with thermoelectric term neglected). The discrepancy between $K$ and $L_{22}$ becomes significant ($18\%$) at the temperature of 40~kK. At $T=100$~kK $L_{22}$ is 3 times larger than $K$.

\begin{figure}
\includegraphics[width=0.95\columnwidth]{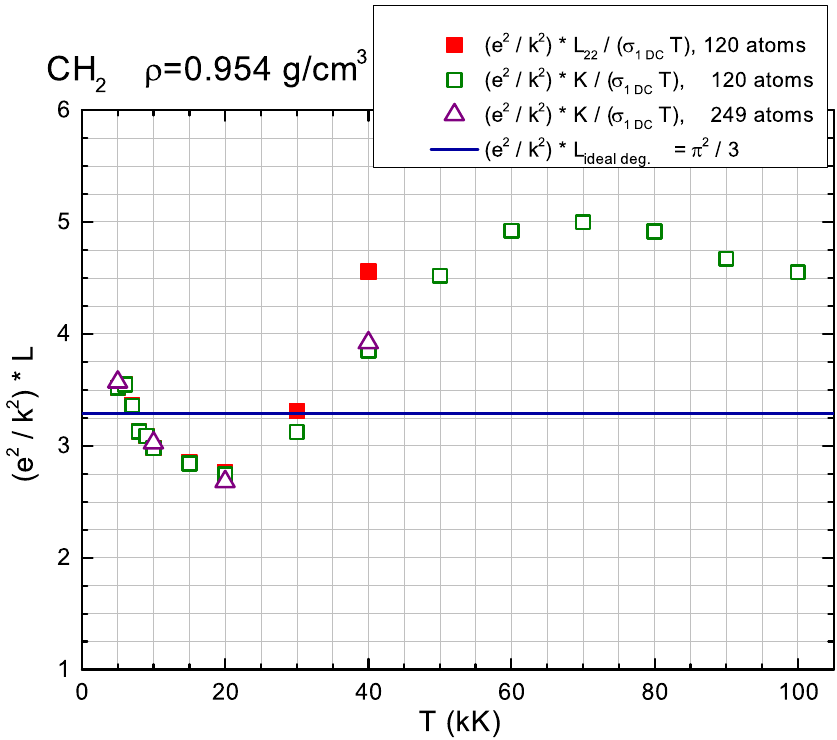}
\caption{Temperature dependences of the calculated Lorenz number. Filled squares---the Lorenz number calculated using the $L_{22}$ Onsager coefficient, 120 atoms. Empty squares---the Lorenz number calculated using thermal conductivity $K$, 120 atoms. Empty triangles---the Lorenz number calculated using thermal conductivity $K$, 249 atoms. Solid line---the ideal degenerate value of the Lorenz number $\frac{e^2}{k^2}L_\mathrm{ideal~deg}=\frac{\pi^2}{3}$.}
\label{Fig:Lorenz}
\end{figure}

Fig.~\ref{Fig:Lorenz} demonstrates the temperature dependences of the calculated Lorenz number $L$; both the Onsager coefficient $L_{22}$ and the thermal conductivity $K$ were used for it. For the sake of convenience, we plot $L$ multiplied by the ratio $e^2/k^2$. The ideal value $\frac{e^2}{k^2}L_\mathrm{ideal~deg}=\frac{\pi^2}{3}$ of the Lorenz number is also plotted. At low temperatures the Lorenz numbers calculated using $L_{22}$ and $K$ evidently coincide. At high temperatures the Lorenz number calculated using the $L_{22}$ coefficient is many times higher than the ideal value.

The Lorenz number calculated using the thermal conductivity $K$ differs from the ideal value; the maximum difference is by a factor of 1.52. The final decision on whether the Wiedemann-Franz law is valid or not depends on the precision required. From the theoretical point of view, the Wiedemann-Franz law for plastics fails. But if the electrical conductivity is known, the thermal conductivity is not available, and the factor of 1.5 may be neglected (common case in hydrodynamic simulations), the Wiedemann-Franz law is a suitable approximation.

\section{DISCUSSION}
\label{Sec:Discussion}

This section contains the brief discussion of the results on the transport properties.

First of all, it should be mentioned that the static electrical conductivity of plastics under the conditions considered has the order of $10^5~\Omega^{-1}$m$^{-1}$. The static electrical conductivity of metals at near-normal densities and temperatures about several tens kK has the order of $10^6~\Omega^{-1}$m$^{-1}$ (see for instance, our previous works \cite{Knyazev:COMMAT:2013, Knyazev:PhysPlasmas:2014}). The static electrical conductivity of metals at the normal conditions usually has the order of $10^7~\Omega^{-1}$m$^{-1}$.

We think, that the most noticeable result on the transport properties obtained in this work is the rapid growth of the static electrical conductivity in the range of temperatures $5\div10$~kK (see Fig.~\ref{Fig:transport}a). To be sure that this is not the computational error caused by the small number of atoms, we have also performed some additional calculations with larger number of atoms in the supercell. The results are presented in Fig.~\ref{Fig:transport}. Neither the results on the electrical conductivity, nor the results on the thermal conductivity have changed qualitatively at increased number of atoms.

To verify our results we have performed comparison with the calculations of other authors at similar conditions. The paper of Horner \textit{et al.} \cite{Horner:PRB:2010} contains the results on the transport and optical properties of plastic of the effective composition CH$_2$ at $\rho=1$~g/cm$^3$ and $T=1$~eV. The calculation is performed by a similar technique. We have performed calculation at the same conditions with the technical parameters described in Sec.~\ref{Sec:Parameters}; the supercell contained 120 atoms.

\begin{figure}
\includegraphics[width=0.95\columnwidth]{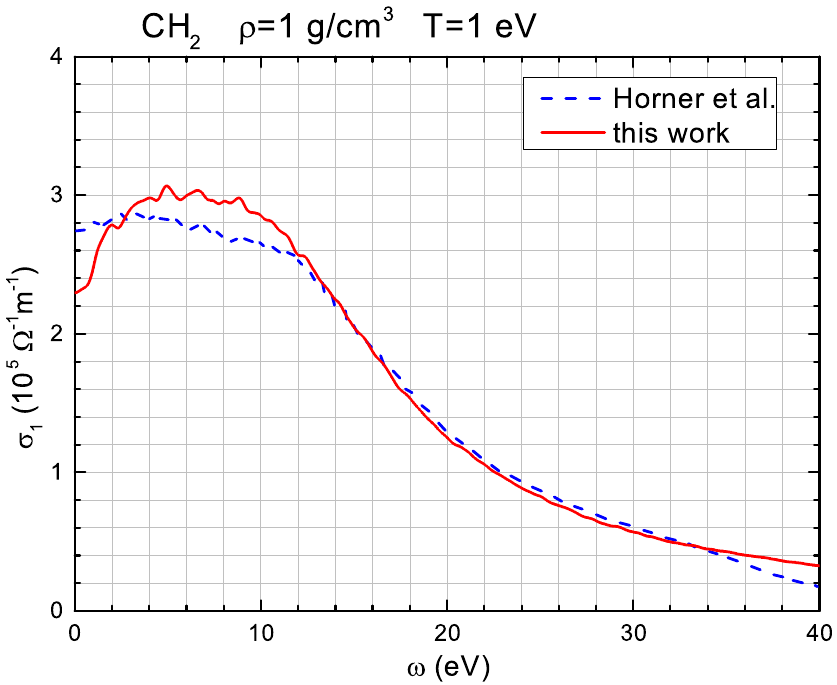}
\caption{The comparison of our results on the dynamic electrical conductivity with those of Horner \textit{et al.}\cite{Horner:PRB:2010}. Solid curve---our results, dashed curve---results of the paper \cite{Horner:PRB:2010}.}
\label{Fig:Horner}
\end{figure}

The comparison of the dynamic electrical conductivity with the paper \cite{Horner:PRB:2010} is shown in Fig.~\ref{Fig:Horner}. The curves are in almost perfect coincidence for the frequencies from 14~eV to 34~eV. There are two regions where the curves diverge: at high frequencies and at low onces.

Our curve has $\sim\omega^{-2}$ asymptotics for the frequencies from 14~eV up to 40~eV; this can be readily checked in the double logarithmic scale. The curve from the work \cite{Horner:PRB:2010} has the $\sim\omega^{-2}$ dependence for the frequencies from 14~eV up to 34~eV. But at higher frequencies the curve of Horner \textit{et al.} suddenly starts to decay much quicker than $\sim\omega^{-2}$. We consider that this rapid decay is due to insufficient number of bands included to the calculation \cite{Horner:PRB:2010} and our curve is more correct for the frequencies higher than 34~eV.

The situation is more complicated at low frequencies. We have used 120 atoms in this calculation, whereas the number of atoms in the calculation of Horner \textit{et al.} was 225. We have checked the influence of the number of atoms on the optical properties for 0.954~g/cm$^3$ isochor; according to our findings it is unlikely that the discrepancy at low frequencies in Fig.~\ref{Fig:Horner} is caused by the different number of atoms. The choice of the initial ionic configuration may be a more possible reason of the discrepancy. At the initial moment we put the separate ions of carbon and hydrogen to the nodes of the auxiliary simple cubic lattice (see Sec.~\ref{Sec:Parameters}). In the paper \cite{Horner:PRB:2010} the CH$_2$ molecules are put to the random positions to form the initial configuration.

The qualitative results of our work coincide with those of the paper \cite{Horner:PRB:2010}. The poor conductivity of the order of $10^5$~$\Omega^{-1}$m$^{-1}$ and the non-Drude shapes of $\sigma_1(\omega)$ with maxima at non-zero frequency were also mentioned in the paper \cite{Horner:PRB:2010}. Our results and the results of Horner \textit{et al.} \cite{Horner:PRB:2010} on $\sigma_{1_\mathrm{DC}}$ at $\rho=1$~g/cm$^3$ and $T=1$~eV are also plotted in Fig.~\ref{Fig:transport}(a).

To learn more about the nature of the static electrical conductivity growth at $\rho=0.954$~g/cm$^3$ (Fig.~\ref{Fig:transport}(a)) we have chosen two temperatures (5~kK and 10~kK) and compared various quantities related to these temperatures.

We have compared the temporal dependences of energy and pressure. Both for 5~kK and 10~kK these dependences are rather common, they are just fluctuating around some equilibrium values.

We can also discuss thermodynamic properties in the temperature range $5\div10$~kK.

The $p-p^\mathrm{kin}_i$ pressure is negative and almost constant ($-114\div-106$~kbar) for 5~kK~$\leq T\leq$~10~kK. The total pressure $p$ is negative at 5~kK ($-23$~kbar) and positive at 10~kK (62~kbar), but this is due to the growing positive kinetic contribution of ions. The heat capacity $C_v$ drops in the temperature range under consideration.

Apparently, some process, that decreases $C_v$, but leaves $p-p^\mathrm{kin}_i$ almost intact, takes place at 5~kK~$\leq T \leq10$~kK. The process, that requires significant amount of energy to be supplied, may cause a sharp maximum on the $C_v(T)$ dependence. In our case we are likely to track the descending branch of such a maximum. This process is also accompanied by the rapid growth of $\sigma_{1_\mathrm{DC}}$.

The ionic configurations at 5~kK and 10~kK were also compared visually for the steps chosen from the equilibrium sections of the QMD simulation. It was found out, that the ions were pretty uniformly distributed over the computational supercell. The absense of the coarse faults (such as unphisical cavities, for intance) was also established.

\begin{figure}
a)~\includegraphics[width=0.95\columnwidth]{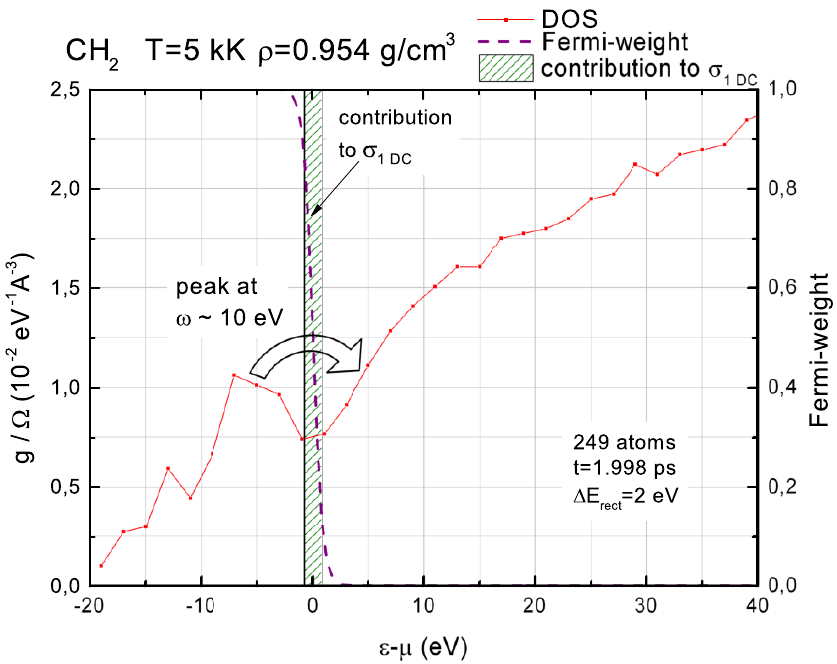}
b)~\includegraphics[width=0.95\columnwidth]{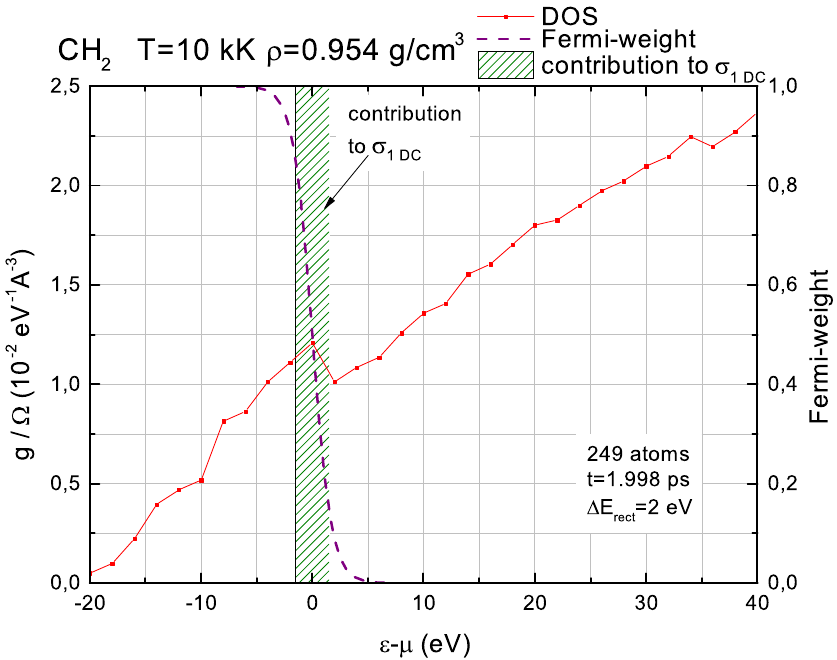}
\caption{The dependence of the electron density of states on the electron energy eigenvalues $\epsilon$ (electron energy for short). Solid curve---electron DOS $g(\epsilon-\mu)/\Omega$ (divided by the volume of the supercell $\Omega$). Dashed line---occupation numbers $f(\epsilon-\mu)=\left(\frac{\epsilon-\mu}{kT}+1\right)^{-1}$, defined by the Fermi-Dirac distribution. Hatching---the $\epsilon-\mu$ values, giving the most significant contribution to $\sigma_{1_\mathrm{DC}}$; determined by FWHM of the $-\frac{\partial f}{\partial\epsilon}(\epsilon-\mu)$ curve. (a) $T=5$~kK. Hatching corresponds to the dip at $g(\epsilon-\mu)$ curve; low $\sigma_{1_\mathrm{DC}}$. The transition yielding the $\omega\approx10$~eV peak at the $\sigma_1(\omega)$ curve (Fig.~\ref{Fig:optical}(a)) is marked by an arrow. (b) $T=10$~kK. Hatching corresponds to rather high $g(\epsilon-\mu)$; high $\sigma_{1_\mathrm{DC}}$.}
\label{Fig:DOS}
\end{figure}

We have also compared the electronic DOS for the temperatures of 5~kK and 10~kK. The results are presented in Fig.~\ref{Fig:DOS}. The density of states $g(\epsilon-\mu)/\Omega$ calculated as described in Sec. \ref{Sec:Technique} is plotted by the solid curve. It should be mentioned that the chemical potential $\mu$ was chosen as the reference point for the electron energies. The DOS was calculated for 249 atoms in the supercell. The occupation numbers $f(\epsilon-\mu)=\left(\frac{\epsilon-\mu}{kT}+1\right)^{-1}$ defined by the Fermi-Dirac distribution are plotted by the dashed curve. The hatching displays the $\epsilon-\mu$ values, which give the most significant contribution to the static electrical conductivity. This hatching corresponds to FWHM (full width at half maximum) of the $-\frac{\partial f}{\partial \epsilon}(\epsilon-\mu)$ curve.

The selection of the hatching area is based on the following reasoning. The Kubo-Greenwood formula may be presented in the following continuous fashion (see, for instance, paper \cite{Collins:PRB:2001}):
\begin{multline}
\sigma_1(\omega)\sim\int\left|\nabla(\epsilon, \epsilon+\hbar\omega)\right|^2\times\\
\times\frac{f(\epsilon)-f(\epsilon+\hbar\omega)}{\hbar\omega}g(\epsilon)g(\epsilon+\hbar\omega)d\epsilon.
\label{Eq:KG_continuous}
\end{multline}
Here $\left|\nabla(\epsilon_1,\epsilon_2)\right|^2$ is some continuous form of the matrix elements $\frac{1}{3}\sum_\alpha\left|\left\langle\Psi_{i}\left|\nabla_\alpha\right|\Psi_{j}\right\rangle\right|^2$. The proportionality sign in (\ref{Eq:KG_continuous}) may contain only fundamental constants and the volume of the supercell $\Omega$. We call Eq.~(\ref{Eq:KG_continuous}) \textit{the continuous Kubo-Greenwood formula}. In the static case $\omega\to0$ Eq.~(\ref{Eq:KG_continuous}) is reduced to:
\begin{equation}
\sigma_{1_\mathrm{DC}}\sim\int\left|\nabla(\epsilon,\epsilon)\right|^2\left(-\frac{\partial f}{\partial \epsilon}(\epsilon)\right)g^2(\epsilon)d\epsilon.
\label{Eq:KG_continuous_DC}
\end{equation}
The contribution of too high or too low $\epsilon$ values is damped by the rapid decay of the $-\frac{\partial f}{\partial \epsilon}(\epsilon)$ curve. So we may estimate the region, yielding the most significant contribution to $\sigma_{1_\mathrm{DC}}$ by the FWHM of the $-\frac{\partial f}{\partial \epsilon}(\epsilon)$ curve. Also we may notice from (\ref{Eq:KG_continuous_DC}), that the higher $g(\epsilon)$ is, the higher is the contribution of the given $\epsilon$ to $\sigma_{1_\mathrm{DC}}$.

The DOS for the temperature 5~kK is shown in Fig.~\ref{Fig:DOS}(a). The area, giving the most significant contribution to the $\sigma_{1_\mathrm{DC}}$ corresponds to the dip at the $g(\epsilon-\mu)$ curve. Thus, the low number of bands contributes to the static electrical conductivity and the values of $\sigma_{1_\mathrm{DC}}$ are relatively low.

Also we may see the peak at the DOS curve at ${\epsilon_1-\mu\approx-5}$~eV.  The bands located at $\epsilon_1-\mu\approx-5$~eV are fully occupied. The transition from these bands to the bands located at $-5$~eV$\lesssim\epsilon_2-\mu\lesssim0$~eV is damped by the high occupation numbers of the final bands. The transition from $\epsilon_1-\mu\approx-5$~eV to 0~eV~$\lesssim\epsilon_2-\mu\lesssim5$~eV is damped by the the low density of states $g(\epsilon_2-\mu)$. The density of the final states $g(\epsilon_2-\mu)$ becomes high enough at $\epsilon_2-\mu\approx5$~eV. Thus, we consider, that the transition from the bands at $\epsilon_1-\mu\approx-5$~eV to $\epsilon_2-\mu\approx5$~eV gives the prominent $\omega\approx10$~eV peak at the curves of the dynamic electrical conductivity (Fig.~\ref{Fig:optical}(a)). This transition is marked by an arrow in Fig.~\ref{Fig:DOS}(a).

The transition from $\epsilon_1-\mu\approx-5$~eV to even higher bands $\epsilon_2-\mu\gtrsim5$~eV is likely damped by the small values of $\left|\nabla(\epsilon_1-\mu,\epsilon_2-\mu)\right|^2$ at high $\epsilon_2-\mu$. This explains the drop of the $\sigma_1(\omega)$ at $\omega\gtrsim10$~eV (Fig.~\ref{Fig:optical}(a)). The $\hbar\omega$ factor in the denominator of Eq.~(\ref{Eq:KG_continuous}) also makes $\sigma_1(\omega)$ drop at high $\omega$.

The DOS for the temperature 10~kK is shown in Fig.~\ref{Fig:DOS}(b). The area, giving the most significant contribution to the $\sigma_{1_\mathrm{DC}}$ now corresponds to the small peak at $g(\epsilon-\mu)$. The values of DOS $g(\epsilon-\mu)$ in the hatched area are larger than those in the hatched area at temperature 5~kK. The larger number of bands contributes to the static electrical conductivity at 10~kK. Consequently, $\sigma_{1_\mathrm{DC}}$ at 10~kK is larger than at 5~kK (Fig.~\ref{Fig:transport}(a)).

The smoother DOS curves at 10~kK also yield the smoother curves of the dynamic electrical conductivity (Fig.~\ref{Fig:optical}(a)).

The bulk of the results on transport properties (Fig.~\ref{Fig:transport}) was obtained for 120 atoms in the supercell by averaging over 15 ionic configurations. Some results (Fig.~\ref{Fig:transport}) were obtained for 249 atoms in the supercell; the averaging was also performed over 15 ionic configurations. It may be readily seen (Fig.~\ref{Fig:transport}) that the qualitative behavior of the results is not affected by the number of atoms. The analysis of the electron DOS is performed for 249 atoms and 1 ionic configuration (corresponding to $t=1.998$~ps). We have also plotted DOS for 120 atoms; the qualitative results have not been affected. The same results were observed when we  tried to choose other ionic configurations. More than one ionic configuration could be used in DOS calculations, however, this would not give anything qualitatively new. So, for the DOS analysis we may use either 120 or 249 atoms and any number of ionic configurations.

In this work we have obtained the step-like behavior of $\sigma_{1_\mathrm{DC}}(T)$ for CH$_2$ plastics along 0.954~g/cm$^3$ isochor. The knee of the $\sigma_{1_\mathrm{DC}}(T)$ dependence is located at approximately 15~kK. Similar behavior for somewhat different conditions was obtained in the number of previous works.

Static electrical conductivity of CH$_2$ plastics along the principal Hugoniot was investigated in the paper of Theofanis \textit{et al.} \cite{Theofanis:PRB:2012} by the wave-packet dynamics method. The step-like behavior of $\sigma_{1_\mathrm{DC}}(T)$ along the principal Hugoniot was obtained. Though the precision of conductivity calculation in the work \cite{Theofanis:PRB:2012} is rather low, it may be established that the knee of $\sigma_{1_\mathrm{DC}}(T)$ dependence is located at approximately 5~kK. The value of conductivity at the plateau in the work \cite{Theofanis:PRB:2012} is $2\div4\cdot10^5~\Omega^{-1}$m$^{-1}$, by the order of magnitude it is close to the value at the plateau in our work. The knee of $\sigma_{1_\mathrm{DC}}(T)$ in the work of Theofanis \textit{et al.} is located at lower temperature than in our work, this is likely because for the same temperatures the density at the principal Hugoniot is higher than normal. The step-like behavior is called "a transition to the metallic state" in the work \cite{Theofanis:PRB:2012}.

The CH plastics along the principal Hugoniot were investigated in the paper \cite{Wang:PhysPlasmas:2011}. The dependences of the static electrical conductivity $\sigma_{1_\mathrm{DC}}(u_s)$ and reflectivity $R(u_s)$ on the shock wave velocity $u_s$ demonstrate a step-like behavior. The CH plastics along the principal Hugoniot were also investigated in the paper \cite{Hu:PRE:2014:1}. The pressure dependence of reflectivity at 532~nm $R(p)$ is reported to have a step-like behavior.

\section{CONCLUSIONS}

\begin{enumerate}
\item In this work we have performed \textit{ab initio} calculation of thermodynamic, transport, and optical properties of plastics of the effective composition CH$_2$ at constant density 0.954~g/cm$^3$; the temperature falls within the range 5~kK~$\leq T\leq100$~kK.

\item The step-like behavior of $\sigma_{1_\mathrm{DC}}(T)$ was obtained. Static electical conductivity grows rapidly for the temperatures 5~kK~$\leq T\leq10$~kK and is almost constant for the temperatures 20~kK~$\leq T\leq60$~kK. 

\item $\sigma_{1_\mathrm{DC}}(T)$ grows rapidly at 5~kK~$\leq T\leq10$~kK simultaneously with the decrease of $C_v(T)$ and the constant behavior of $p(T)-p^\mathrm{kin}_i(T)$.

\item The behavior of $\sigma_{1_\mathrm{DC}}(T)$ was partially explained by the investigation of the electron density of states. The rapid growth of $\sigma_{1_\mathrm{DC}}(T)$ at 5~kK~$\leq T\leq10$~kK is connected with the growth of the DOS at $\epsilon=\mu$.

\end{enumerate}

\section*{ACKNOWLEDGEMENTS}

We thank Yu. K. Kurilenkov for stimulating discussion on the electron DOS.

This work was supported by the FAIR-Russia Research Center Grants (2014--2015), the Russian Foundation for Basic Research, Grant Nos. 14-08-31450, 13-08-01179	and 13-02-91057. We acknowledge the project of the Ministry of Education and Science of the Russian Federation, project No. 3.522.2014/K.


%

\end{document}